\pdfoutput=1 
\documentclass[aps,twocolumn,prl,showpacs,amsmath,amssymb,10pt]{revtex4-1}
\usepackage{graphicx}
\usepackage{scrextend}
\usepackage{manfnt,pifont}

\usepackage{url}

\begin{document}


\title{Loop Corrections to Supergravity on $AdS_5 \times S^5$}

\author{Luis F. Alday$^{a}$}
\author{Agnese Bissi$^{b}$}
\affiliation{\it $^{a}$Mathematical Institute, University of Oxford,Andrew Wiles Building, Radcliffe Observatory Quarter,Woodstock Road, Oxford, OX2 6GG, UK\\
\it $^{b}$Center for the Fundamental Laws of Nature, Harvard University, Cambridge, MA 02138 USA}

\date{\today}

\begin{abstract}

\noindent
We consider the four-point correlator of the stress-energy tensor multiplet in ${\cal N}=4$ SYM. In the planar limit and at large 't Hooft coupling such correlator is given by the corresponding holographic correlation function in IIB supergravity on $AdS_5 \times S^5$. We consider subleading corrections in the number of colours, {\it i.e.} order $1/N^4$, at large 't Hooft coupling. This corresponds to loop corrections to the supergravity result. Consistency conditions, most notably crossing symmetry, constrain the form of such corrections and lead to a complete determination of the spectrum of leading twist intermediate operators.  

\end{abstract}

\pacs{11.15.Pg, 11.25.Hf, 11.25.Tq}


\maketitle

\noindent
{\bf Introduction.}  The prototypical example of $AdS/CFT$ correspondence relates ${\cal N}=4$ Super Yang-Mills (SYM) in four dimensions to type IIB string theory on $AdS_5 \times S^5$ \cite{Maldacena:1997re,Gubser:1998bc,Witten:1998qj}. Even after 20 years of its original formulation, and in spite of tremendous progress in many directions, non-protected quantities have only been explored in certain corners, or edges, of the parameter space. One particularly interesting corner corresponds to planar SYM with large t'Hooft coupling $\lambda=g_{YM}^2 N$, which is dual to classical supergravity on the bulk. In this regime single trace chiral primary operators (CPO) of weight $p$, ${\cal O}_p$,  map to supergravity fields with mass $m^2=p(p-4)$, and their correlation functions can in principle be computed by tree-level Witten diagrams on $AdS$:
\begin{equation}
\langle {\cal O} \cdots {\cal O} \rangle_{conn.} \sim \frac{1}{N^2} \left( \text{tree-level Witten diagram}\right)
\end{equation}
Three point correlators of arbitrary CPOs, as well as four point correlator of the stress tensor multiplet, were computed long ago \cite{Freedman:1998tz, DHoker:1998vkc}. Recently, an elegant algorithm based on symmetries and consistency conditions to determine the four-point correlator of arbitrary CPOs was proposed in \cite{Rastelli:2016nze}, see also \cite{Dolan:2006ec}.     

In this note we consider subleading corrections in $1/N$ to correlators in the large t' Hooft coupling regime. This corresponds to quantum corrections on the gravity side. Although some progress has been made for specific contributions, see \cite{Fitzpatrick:2011hu}, loop diagrams in $AdS$ are a largely unexplored subject, mostly due to technical difficulties that prohibit direct computations. The analytic bootstrap was initiated in  \cite{Komargodski:2012ek,Fitzpatrick:2012yx} and developed into a powerful algebraic machinery in \cite{Alday:2014tsa,Alday:2015ewa,Alday:2016njk}. This algebraic formulation allowed a systematic study of loops in $AdS$ based on symmetries, started in \cite{Aharony:2016dwx} for sectors of CFTs. In this letter we would like to report the first complete results to order $1/N^4$ for a full fledge CFT, namely ${\cal N}=4$ SYM. We will focus on the four point correlator of the lowest component of the stress tensor multiplet ${\cal O}_2$. In the planar limit and at large 't Hooft coupling the intermediate operators consist of double trace operators of spin $\ell$ and dimension
\begin{equation*}
\Delta_{n,\ell} = 4+2n+\ell+ \frac{\gamma_{n,\ell}^{(1)}}{N^2} +\frac{\gamma_{n,\ell}^{(2)}}{N^4} + \cdots
\end{equation*}
where $n=0,1,\cdots$. The leading order correction $\gamma_{n,\ell}^{(1)}$ is given by the supergravity result. In this letter we study the consequences of superconformal symmetry, consistency of the OPE and crossing symmetry for the subleading corrections. The analysis of crossing symmetry to order $1/N^4$ is highly complicated by mixing among double trace operators. Namely, there is more than one intermediate operator for a given twist and spin. From the bulk point of view this corresponds to take into account all Kaluza Klein (KK)-modes. In order to solve this mixing problem we study general correlators $\langle {\cal O}_p{\cal O}_p{\cal O}_q{\cal O}_q \rangle$ in the supergravity approximation. This allows to disentangle the contribution of each double trace operator/KK-mode and apply the methods of \cite{Alday:2016njk,Aharony:2016dwx} for the present case. This leads to an expression for $\gamma_{n,\ell}^{(2)}$ valid to all orders in inverse powers of the spin for each KK-mode. These expansions can be resummed exactly. 

In addition crossing symmetry allows the addition of solutions with finite support in the spin. From the bulk perspective these ambiguities correspond to unknown coefficients in front of possible counterterms. For the present case we expect such extra solutions to be absent for spin two and higher. With this, for instance, for the leading twist operators of spin two and four we obtain
\begin{eqnarray*}
\Delta_{0,2}&=& 6 - \frac{4}{N^2}- \frac{45}{N^4} + \cdots\\
\Delta_{0,4}&=& 8 - \frac{48}{25}\frac{1}{N^2}-\frac{12768}{3125}\frac{1}{N^4}+\cdots
\end{eqnarray*}
Similar results can be obtained for any spin. In principle our algorithm fixes also the OPE coefficients.  

{\bf Note added:} Shortly after our paper appeared in arXiv, an independent computation was presented \cite{Aprile:2017bgs}. Our results appear to be in full agreement. 

\bigskip

\noindent
{\bf Stress-tensor correlator in ${\cal N}=4$ SYM.} In ${\cal N}=4$ SYM the stress tensor sits in a half-BPS multiplet. The lowest component of this multiplet is a scalar operator ${\cal O}_2$ of dimension two transforming in the $[0,2,0]$ of the $R-$symmetry group $SU(4)$. Its four-point correlator takes the form
\begin{equation}
\nonumber
\langle {\cal O}_{2}(x_1) {\cal O}_{2}(x_2) {\cal O}_{2}(x_3) {\cal O}_{2}(x_4)\rangle = \sum_{\cal R} \frac{{\cal G}^{({\cal R})}(u,v)}{x_{12}^4x_{34}^4}
\end{equation}
where the sum runs over representations in the tensor product $[0,2,0] \times [0,2,0]$ and we have introduced the standard cross-ratios
\begin{equation*}
u= \frac{x_{12}^2 x_{34}^2}{x_{13}^2 x_{24}^2},~~~v= \frac{x_{14}^2 x_{23}^2}{x_{13}^2 x_{24}^2}
\end{equation*}
Superconformal symmetry allows writing all contributions ${\cal G}^{({\cal R})}(u,v)$ in terms of a single non-protected function ${\cal G}(u,v)$ satisfying the following crossing relation. 
\begin{equation*}
v^2 {\cal G}(u,v) -u^2 {\cal G}(v,u) + 4(u^2-v^2) + \frac{4(u-v)}{c}=0
\end{equation*}
where $c=\frac{N^2-1}{4}$ is the central charge. See \cite{Beem:2013qxa,Beem:2016wfs} for a detailed discussion. This function can be decomposed into the contribution from operators in (semi-)short multiplets and operators in long multiplets
\begin{equation*}
{\cal G}(u,v)={\cal G}^{short}(u,v)+ \mathcal{H}(u,v)
\end{equation*}
where ${\cal G}^{short}(u,v)$ is independent of the coupling constant and can be found in \cite{Beem:2016wfs}, and $\mathcal{H}(u,v)$ admits a decomposition in superconformal blocks
\begin{equation*}
\mathcal{H}(u,v) = \sum_{\tau,\ell} a_{\tau,\ell} u^{\tau/2} g_{\tau+4,\ell}(u,v)
\end{equation*}
where the sum runs over superconformal primary operators in long multiplets, in the singlet representation of $SU(4)$, with twist (dimension minus the spin) $\tau$ and even spin $\ell$. $a_{\tau,\ell}$ denotes the square of the OPE coefficients. It is convenient to write the conformal blocks in terms of cross-ratios $(z,\bar z)$ such that $z \bar z=u,(1-z)(1-\bar z)=v$. In terms of these
\begin{equation*}
g_{\tau,\ell}(z,\bar z) = \frac{z^{\ell+1} F_{\frac{\tau}{2}+\ell}(z) F_{\frac{\tau-2}{2}}(\bar z) - \bar z^{\ell+1} F_{\frac{\tau}{2}+\ell}(\bar z) F_{\frac{\tau-2}{2}}(z) }{z-\bar z}
\end{equation*}
where $F_\beta(z)=\,_2F_1(\beta,\beta,2\beta;z)$ is the standard hypergeometric function.
In the strict limit of infinite central charge $\mathcal{H}(u,v)$ reduces to the generalised free fields result $\mathcal{H}^{(0)}(u,v)$, which agrees with the large $c$ result in the Born approximation (free theory). The intermediate operators correspond to towers of double trace operators of twist $\tau_n=4+2n$ and OPE coefficients 
\begin{equation*}
a^{(0)}_{n,\ell}=\frac{\pi (\ell+1)  (\ell+2 n+6) \Gamma (n+3) \Gamma (\ell+n+4)}{2^{2 \ell+4 n+9} \Gamma \left(n+\frac{5}{2}\right) \Gamma \left(\ell+n+\frac{7}{2}\right)}
\end{equation*}
The four-point correlator admits an expansion around large $N$ or equivalently large central charge $c$:
\begin{equation*}
\mathcal{H}(u,v) = \mathcal{H}^{(0)}(u,v) + \frac{1}{c} \mathcal{H}^{(1)}(u,v)+ \frac{1}{c^2} \mathcal{H}^{(2)}(u,v)+\cdots
\end{equation*}
Accordingly the intermediate operators acquire corrections
\begin{eqnarray}
\label{dataexpansion}
\tau_{n,\ell} &=& 4+2n + \frac{1}{c} \gamma^{(1)}_{n,\ell} + \frac{1}{c^2} \gamma^{(2)}_{n,\ell} + \cdots\\
a_{n,\ell}&=&a^{(0)}_{n,\ell}+\frac{1}{c} a^{(1)}_{n,\ell}+\frac{1}{c^2} a^{(2)}_{n,\ell}+\cdots \nonumber
\end{eqnarray}
In this note we will focus in the limit of large 't Hooft coupling $\lambda$. In this regime there is no new operators appearing in the OPE at this order and $\mathcal{H}^{(1)}(u,v)$ can be computed from the classical supergravity result in \cite{Dolan:2006ec,Rastelli:2016nze}. This leads to the following correction for the spectrum and OPE coefficients \cite{DHoker:1999mic, Arutyunov:2000ku, Alday:2014tsa}
\begin{eqnarray*}
 \gamma^{(1)}_{n,\ell} &=& -\frac{\kappa_n}{(1+\ell)(6+\ell+2n)},\\
 a_{n,\ell}^{(1)} &=& \frac{1}{2} \partial_n\left(  a_{n,\ell}^{(0)}  \gamma^{(1)}_{n,\ell} \right)
\end{eqnarray*}
where $\kappa_n=(n+1)(n+2)(n+3)(n+4)$. It is important to note that for a given $n$ and $\ell$ there is more than one superconformal primary in the singlet of $SU(4)$, except for $n=0$. The above corrections should then be interpreted as (weighted-)averages.  This will be very important below. 

\bigskip

\noindent
{\bf From leading to subleading corrections.} Our aim is to compute $\gamma^{(2)}_{n,\ell}$ and $a^{(2)}_{n,\ell}$ from crossing symmetry. Since ${\cal G}^{short}(u,v)$ receives contributions only up to order $1/c$, the crossing equation for $\mathcal{H}^{(2)}(u,v)$ is simply
\begin{equation}
\label{crossing}
v^2 \mathcal{H}^{(2)}(u,v) = u^2 \mathcal{H}^{(2)}(v,u).
\end{equation}
We will follow the same strategy as in \cite{Aharony:2016dwx}: start by determining the piece proportional to $\log^2u$ in $\mathcal{H}^{(2)}(u,v)$ from the CFT data at order $1/c$. By crossing symmetry this will lead to a precise divergence proportional to $\log^2 v$. Matching this divergence then fixes  $\gamma^{(2)}_{n,\ell}$ and  $a^{(2)}_{n,\ell}$ to all orders in $1/\ell$. Plugging (\ref{dataexpansion}) into the conformal block decomposition and expanding up to order $1/c^2$, we find
\begin{align}
\label{H2CPW}
\mathcal{H}^{(2)}(u,v) = &\sum_{n,\ell}  \Bigg(a^{(2)}_{n,\ell} + \frac{1}{2} a^{(0)}_{n,\ell} \gamma^{(2)}_{n,\ell} \partial_n  
 + \frac{1}{2}a^{(1)}_{n,\ell} \gamma^{(1)}_{n,\ell} \partial_n \cr
 &+ \frac{1}{8}a^{(0)}_{n,\ell}  (\gamma^{(1)}_{n,\ell} )^2  \partial^2_n \Bigg) u^{2+n} g_{n,\ell}(u,v)\,
\end{align}
where we have introduced $g_{n,\ell}(u,v) \equiv g_{\tau_{n}^{(0)}+4,\ell}(u,v)$. In particular the piece proportional to $\log^2u$ is
\begin{align}
\label{H2log}
\left. \mathcal{H}^{(2)}(u,v) \right|_{\log^2u} = \sum_{n,\ell}  \frac{1}{8}a^{(0)}_{n,\ell}  (\gamma^{(1)}_{n,\ell} )^2  u^{2+n} g_{n,\ell}(u,v).
\end{align}
A serious obstacle to compute this is the mixing among double trace operators $[{\cal O}_2,{\cal O}_2]_{n,\ell}, [{\cal O}_3,{\cal O}_3]_{n-1,\ell}, \dots$. They have the same twist and spin at zeroth order and transform under the same representation of $SU(4)$. Hence, the sum in (\ref{H2log}) should contain an extra index $I$ - which we leave implicit- to account for degenerate operators at tree-level. For the same reason, quantities above should be interpreted as averages over these families, weighted by their respective OPE coefficient at zeroth order. Therefore, the weighted average $\langle (\gamma^{(1)}_{n,\ell} )^2\rangle$ does not follow from the leading order result, except for $n=0$, for which there is a unique state. In order to tackle this mixing problem we consider the complete families of four-point correlators $\langle {\cal O}_p {\cal O}_p{\cal O}_q{\cal O}_q\rangle$ in the supergravity approximation. This is done in the appendix and it leads to the following remarkable structure for the average in question
\begin{equation}
\label{gsexp}
\langle  (\gamma^{(1)}_{n,\ell} )^2 \rangle= \frac{ \kappa_n^3 (5+2n)}{120(J^2-(n+2)(n+3))^2} + \sum_{j=2}^{n+2} \frac{\beta_{n,j}}{J^2-j(j+1)}
\end{equation}
where $J^2=(\ell+n+3)(\ell+n+4)$. The coefficients $\beta_{n,j}$ have been computed explicitly up to twist 10. We will see however that they can be fixed, for any value of the twist, by resorting to crossing symmetry. In order to understand this, consider the following sequence of twist conformal blocks (TCB)
\begin{equation*}
H^{(m)}_n(z,\bar z)= \sum_{\ell} \frac{a^{(0)}_{n,\ell}}{J^{2m}} u^{n}  g_{n,\ell}(z,\bar z)
\end{equation*}
We then propose the following expansion:
\begin{equation*}
\langle (\gamma^{(1)}_{n,\ell} )^2\rangle= \sum_m \frac{c_m^n}{J^{2m}}
\end{equation*}
This allows to write the piece proportional to $\log^2 u$ in $\mathcal{H}^{(2)}(u,v)$ in terms of TCB and the coefficients $c_m^n$. Furthermore, crossing plus consistency with the CPW expansion, {\it e.g.} absence of $\log^3 v$, fixes the range of $m$ to be $m=2,3,\cdots$. From the explicit expression  for TCB found in the appendix, we can extract the contribution proportional to $\log^2 v$.  
\begin{align*}
\left. \mathcal{H}^{(2)}(u,v) \right|_{\log^2u \log^2 v}= \frac{1}{8} \sum_{m,n}c_m^n \frac{u^{n+2}}{\bar z-z} F_{n+3}(z) q_n^{(m)}(\bar z)
\end{align*}
where the functions $q_n^{(m)}(\bar z)$ are defined in the appendix. This contribution should be crossing symmetric by itself. This imposes a set of linear constraints on the coefficients $c_m^n$. Now we note the following remarkable fact: the expansion (\ref{gsexp}) is consistent with this set of constraints and furthermore, the constraints fix uniquely the coefficients $\beta_{n,j}$ for all twists! Up to twist 10 these coefficients agree precisely with the ones found by explicit computations. 

Having found the averages $\langle (\gamma^{(1)}_{n,\ell} )^2\rangle$ we can now turn into the sums
\begin{equation}
\label{Sn}
S_n(z,\bar z) \equiv \sum_\ell  a^{(0)}_{n,\ell}  \left(\gamma^{(1)}_{n,\ell}\right)^2 u^n g_{n,\ell}(z,\bar z)
\end{equation}
These sums can be decomposed into individual contributions, corresponding to the insertion of single or double poles in $J^2$ into the sums defining the TCB. Denoting by ${\cal C}$ the quadratic Casimir with eigenfunction $u^n g_{n,\ell}(z,\bar z)$ and eigenvalue $J^2$ we obtain
\begin{equation}
\nonumber
({\cal C}-j (j+1))\left( \sum_\ell   \frac{a^{(0)}_{n,\ell} }{J^2-j(j+1)} u^n g_{n,\ell}(z,\bar z)\right) = H^{(0)}_n(z,\bar z)
\end{equation}
which gives a differential equation for the components of the sums (\ref{Sn}). This can be easily solved case by case. The sums have the following structure
\begin{equation}
\label{Sstructure}
S_n(z,\bar z) = \frac{u^n}{z-\bar z}\left(F_{n+3}(\bar z) s_n(z)-F_{n+3}( z) s_n(\bar z) \right)
\end{equation}
for instance, for the first case we obtain
\begin{equation}
\nonumber
s_0(z)=\frac{48 \log (1-z) ((z^2-6z+6) \log (1-z)-3 z^2+6z)}{z^5}
\end{equation}

\bigskip

\noindent
{\bf Spectrum at order $1/c^2$.} We will now consider the crossing equation (\ref{crossing}). Our strategy will be to expand it around $z=0,\bar z=1$ and focus in terms proportional to different powers of $\log z$ and $\log(1-\bar z)$. Note that the $\log z$  dependence in (\ref{H2CPW}) will only arise when the derivative hits $u^{2+n}$. On the other hand the behaviour around $\bar z=1$ is more subtle and one needs to perform the sum over the spin. The piece proportional to $\log^2z \log^2(1-\bar z)$ has already been discussed in the previous section regarding the problem of mixing. The relation proportional to $\log z \log^2(1-\bar z)$ leads to
\begin{align*}
&\sum_{n,\ell} u^{n}\left( \frac{1}{2}\left( a^{(0)}_{n,\ell} \gamma^{(2)}_{n,\ell} + a^{(1)}_{n,\ell} \gamma^{(1)}_{n,\ell}\right) +a^{(0)}_{n,\ell} (\gamma^{(1)}_{n,\ell})^2 \frac{\partial_n}{4} \right) g_{n,\ell}(z,\bar z) \\
&+\sum_n \frac{\log \bar z}{4} S_n(z,\bar z) \Bigg|_{\log^2(1-\bar z)} = \left. \frac{1}{8} \sum_n S_n(1-\bar z,1-z) \right|_{\log z}
\end{align*}
In this note we will restrict ourselves to corrections to the spectrum of leading twist operators $\gamma^{(2)}_{0,\ell}$. This amounts to consider the small $z$ limit of the relation above. On the l.h.s. only terms with $n=0$ will survive. On the other hand, note that all terms on the r.h.s will contribute to this limit. The sum over derivatives of conformal blocks with the extra insertion $(\gamma^{(1)}_{n,\ell})^2$ can be computed with some effort. With this result together with the derivative relation for $a^{(1)}_{n,\ell}$ the above relation in the small $z$ limit can be expressed as follows
\begin{eqnarray}
\sum_{\ell} \frac{1}{2} a^{(0)}_{0,\ell} \hat \gamma^{(2)}_{0,\ell} g^{coll}_{0,\ell}(\bar z) + \frac{\partial_n}{4} \left. \left( \kappa_n^2 \rho_n F_{3+n}(\bar z) \log(1-\bar z) \right) \right|_{n=0} \nonumber \\
+ \frac{\log \bar z}{4} S_0(z,\bar z) \Bigg|_{z^0 \log^2(1-\bar z)} = \left. \frac{1}{8} \sum_n S_n(1-\bar z,1-z) \right|_{z^0 \log z} \label{crossing12}
\end{eqnarray}
where $g^{coll}_{0,\ell}(\bar z)$ is the small $z$ limit of $g_{0,\ell}(z,\bar z)$ and we have introduced $\hat \gamma^{(2)}_{n,\ell}=\gamma^{(2)}_{n,\ell}-1/2 \gamma^{(1)}_{n,\ell} \partial_n \gamma^{(1)}_{n,\ell}$. All the terms except the first one in the above relation are exactly computable. Crossing symmetry then implies that $\hat \gamma^{(2)}_{0,\ell}$ should be such that its insertion produces a $\log^2(1-\bar z)$ divergence times a fully fixed expansion in powers of $(1-\bar z)$. This problem can be solved by proposing the following expansion
\begin{equation*}
\hat \gamma^{(2)}_{0,\ell}= \sum_m \frac{b_m}{J^{2m}}
\end{equation*}
Hence the first term in (\ref{crossing12}) can be written in terms of TCB $H^{(m)}_n(z,\bar z)$ at $z=0$. As before, crossing symmetry plus consistency with the CPW expansion fixes the range $m=2,3,\cdots$. From the procedure outlined in the appendix one can compute the contribution proportional to $\log^2(1-\bar z)$ for $h^{(m)}_n(\bar z)$ for $m=2,3,\cdots$. This allows to determine all coefficients $b_m$, and hence $\hat \gamma^{(2)}_{0,\ell}$ to all orders in $1/\ell$. The result can be organised as to make manifest the contribution from each KK-mode. We start by representing $\langle (\gamma^{(1)}_{n,\ell} )^2\rangle$ as follows 
\begin{equation*}
\langle (\gamma^{(1)}_{n,\ell} )^2\rangle =\sum_{p=2}^\infty \prod_{k=2}^{p-1} \frac{\alpha_p \kappa_n^2 (n-k+1)(n+k+3)}{(J^2-(n+2)(n+3))^2(J^2-k(k+2))}
\end{equation*}
where $\alpha_p=p^2(p^2-1)/12$. Each term inside the sum represents the contribution from the $p-$th KK mode, or more precisely the intermediate double trace operators $[{\cal O}_p,{\cal O}_p]$. We can then compute the contribution to $\hat \gamma^{(2)}_{0,\ell}$ from each KK-mode. From the bulk point of view, this has the interpretation of an expansion into KK-modes running along the loop. Following the steps outlined above, we can compute $\hat \gamma^{(2)}_{0,\ell}$ to all orders in $1/\ell$. Remarkably, the resulting series can be resummed exactly. For the massless KK-modes one obtains
\begin{equation*}
\left. \hat \gamma^{(2)}_{0,\ell} \right|_{p=2} = -\frac{144(3 J^4-2J^2+4)}{(J^2-6)^2(J^2-2)J^2}
\end{equation*}
Taking into account only the massless KK-mode should be equivalent to doing the bulk computation in 5d super-gravity. Note that in this case the answer is convergent and finite for all values of the spin. This is consistent with the fact that 5d supergravity is free of divergences at one loop. For $p=3,4,\cdots$ the results have the following structure
\begin{equation*}
\left. \hat \gamma^{(2)}_{0,\ell} \right|_{p} = \frac{P^{(2p+6)}(\ell)}{J^2(J^2-2)(J^2-6)^2} + \frac{Q^{(p+1)}(J^2)}{J^2-6}\psi^{(2)}(\ell+1)
\end{equation*}
where $P$ and $Q$ are polynomials such that this contribution starts at order $J^{-2p}$ at large $J$. An important comment is in order. Even though the contribution of each KK-mode leads to an asymptotic series in $1/J$, the sum of all of them leads to a convergent series. This is in tune with the analysis of \cite{Caron-Huot:2017vep}. Let us now consider the contribution of a generic KK-mode for finite/small values of the spins. The general structure is
\begin{eqnarray}
\nonumber
\left. \hat \gamma^{(2)}_{0,\ell} \right|_{p}& =&\alpha_p \frac{P^{(14+2\ell)}(p)}{(p^2-4)(p^2-1)p}\\
&&+\alpha_p (p^2-4)(p^2-1)p^3Q^{(4+2\ell)}(p) \psi^{(2)}(p) \nonumber
\end{eqnarray}
for some polynomials $P,Q$.  At large $p$ we find the following behaviour
\begin{eqnarray*}
\nonumber
\left. \hat \gamma^{(2)}_{0,\ell} \right|_{p} \sim \frac{\alpha_p}{p^{3+2\ell}}
\end{eqnarray*}
since $\alpha_p \sim p^4$, this implies the sum over $p$ is actually divergent for spin zero! Note that this agrees with the presence of a quadratic divergence in the 10d supergravity computation. For spin two and higher we get a convergent sum. For instance, for the first cases we obtain
\begin{eqnarray*}
\nonumber
\sum_{p=3} \left. \hat \gamma^{(2)}_{0,2} \right|_{p} =-\frac{4523}{1680}\\
\sum_{p=3} \left. \hat \gamma^{(2)}_{0,4} \right|_{p} =-\frac{3832}{21875}
\end{eqnarray*}
which leads to the results quoted in the introduction. Similar results are obtained for arbitrary spin \cite{workinprogress}. 

\bigskip

\noindent
{\bf Discussion.} In this note we have reported the first complete results for the CFT data of unprotected operators in ${\cal N}=4$ SYM to order $1/N^4$ and at large 't Hooft coupling. A more detailed exposition will appear in \cite{workinprogress}. There are several open questions that would be nice to adress. 

It would be interesting to compute explicitly $ \gamma^{(2)}_{n,\ell}$ for $n>0$. Once this is found it would be interesting to study its large $n$ behaviour and compare it to the expectations from the bulk perspective. It would be important to understand if solutions with finite support in the spin are present. Preliminary results show that crossing symmetry does not require any non-analytical corrections at finite spin. On the other hand crossing symmetry allows the addition of any of the truncated solutions constructed in  \cite{Alday:2014tsa,Heemskerk:2009pn}. From the bulk perspective these  solutions correspond to counterterms, {\it e.g.} the ones that need to be added to render the computation finite. The 10d supergravity computation contains a quadratic divergence proportional to $\lambda^{1/2} {\cal R}^4$, see \cite{Banks:1998nr} eq. 4.2. This will lead to a contribution which becomes large for large $\lambda$ but has support only for spin zero. Indeed, this divergence appears to be visible in our computation, when summing over KK-modes for spin zero. We expect that other extra solutions are not present. Note that this ambiguity is already present at leading order in $1/c$. In this case, all truncated solutions are forbidden by requiring consistency with the flat space limit, see {\it e.g.}\cite{Rastelli:2016nze}. Presumably consistency with the flat space amplitude  to order $1/c^2$ will also forbid most extra solutions. Relatedly, there are several results in the literature that bound the behaviour of $\gamma_{n,\ell}^{(1)}$ for large $n$, see  {\it e.g.} \cite{Maldacena:2015iua, Cornalba:2006xm, Cornalba:2007zb, Kulaxizi:2017ixa,Li:2017lmh}, it would be interesting to extend these results to the order we are considering. 

Leading order corrections in $1/\lambda$ are in principle possible to consider. At leading order they correspond to the addition of the first truncated solution with a known coefficient \cite{Goncalves:2014ffa}. At order $1/N^4$ one would have to 'square' the supergravity contribution plus this contribution. Since the latter is truncated in the spin, the extra sums involved are very simple. This computation is also expected to lead to divergences, since the first truncated solution grows much faster, with $n$, than supergravity. One could also consider the exchange of a finite number of single trace operators, combining the results of \cite{Alday:2017gde} with the methods of this note. 

It would be interesting to study the full four-point correlator in space time. In this note we have computed explicitly the piece proportional to $\log^2u$, which should encode the full physical information about the correlator \cite{Caron-Huot:2017vep}. For instance, we have seen that from this piece, through crossing, the CFT data follows to all orders in $1/\ell$, and from this the four point correlator can be reconstructed, up to pieces which contribute only for finite values of the spin given in \cite{Alday:2014tsa}.  It would be interesting to study this problem in Mellin space. This would be the first step to extend the results of \cite{Rastelli:2016nze} to include loop corrections. 

The expansion in $1/N$ for non-protected quantities in the context of $AdS/CFT$ duality is a largely unexplored subject. Our result opens a window to study this problem systematically and quantitatively.  

\bigskip

\begin{acknowledgments}
We are grateful to G. Arutyunov, Z. Komargodski, J. Maldacena, A. Zhiboedov and specially Ofer Aharony for useful discussions. We would like to thank the ICTP-SAIFR in Sao Paulo for their hospitality during part of this work. The work of L.F.A was supported by ERC STG grant 306260. L.F.A. is a Wolfson Royal Society Research Merit Award holder. The work of A.B. is partially supported by Templeton Award 52476 of A. Strominger and by Simons Investigator Award from the Simons Foundation of X. Yin.  
\end{acknowledgments}

\appendix

\section{Appendices}

\noindent
{\bf The Mixing problem}. A technical obstacle in inferring the $\log^2 u$ piece of the correlator at order $1/c^2$ is mixing. For a given twist $\tau_n=4+2n$ all double trace operators $[{\cal O}_2,{\cal O}_2]_{n},[{\cal O}_3,{\cal O}_3]_{n-1}, \cdots$ mix, and the eigenfunctions of the Hamiltonian are certain combinations of those
\begin{equation*}
\Sigma_i = \alpha^{2}_i [O_2,O_2]_{n} + \cdots +\alpha^n_i [O_{2+n},O_{2+n}]_{0} 
\end{equation*}
where the dependence on the spin is implicit. We will consider this problem at leading order in $1/c$. We can choose the double trace operators $[{\cal O}_k,{\cal O}_k]$ to be canonically normalised, such that the coefficients $\alpha^{p}_i$ form an orthonormal matrix in this basis. In order to solve the mixing problem we consider general correlators $\langle {\cal O}_p{\cal O}_p{\cal O}_q{\cal O}_q \rangle$. At zeroth order the above operators appear in this correlator with OPE coefficient
\begin{equation*}
\sum_i c_{pp\Sigma_i}c_{qq\Sigma_i}= \eta_p \eta_q \sum_i \alpha_i^p \alpha_i^q
\end{equation*}
where $\eta_{p},\eta_{q}$ could also depend on the spin and the twist. Note that this sum is proportional to $\delta_{pq}$. At order $1/c$ these operators acquire an anomalous dimension $\gamma_{i}$. From the explicit conformal block decomposition for the correlator $\langle {\cal O}_p{\cal O}_p{\cal O}_q{\cal O}_q \rangle$ in the supergravity approximation we can read off 
\begin{equation*}
\sum c_{pp\Sigma_i}c_{qq\Sigma_i} \gamma_i= \eta_p \eta_q \sum_i \alpha_i^p \alpha_i^q \gamma_i \equiv \eta_p \eta_q  \langle \gamma \rangle_{pq}
\end{equation*}
the averages $\langle \gamma \rangle_{pq}$ for a given twist can be conveniently packed in a mixing matrix $M^{(n)}$. For instance, for $n=1$ we have to analyse the correlators with $p,q=2,3$, which can be found in \cite{Dolan:2004iy}. This leads to the following mixing matrix
\begin{equation*} 
M^{(1)}= \kappa_1 \left(
\begin{array}{cc}
 -\frac{1}{J^2-12} & -\frac{6}{(J^2-12) \sqrt{J^2-6}} \\
 -\frac{6}{(J^2-12) \sqrt{J^2-6}} & -\frac{J^2+24}{J^4-18 J^2+72} \\
\end{array}
\right)
\end{equation*}
where $J^2=(\ell+4)(\ell+5)$ for $n=1$. We have analysed this problem for several values of $p,q$, using the explicit supergravity results in \cite{Dolan:2004iy, Uruchurtu:2011wh, Arutyunov:2003ae}. The averages $ \langle \gamma^2 \rangle_{pq}$ at order $1/c^2$  are given by the elements of $M^{(n)}\cdot M^{(n)}$. We are interested in $ \langle \gamma^2 \rangle_{22}$. For instance, for the case $n=1$ we obtain
\begin{equation*}
 \langle \gamma^2_{n=1,\ell} \rangle_{22} = \kappa_1^2\left( \frac{7}{(J^2-12)^2} + \frac{1}{J^2-6}-\frac{1}{J^2-12} \right)
\end{equation*}
 In all cases we found the remarkable pattern (\ref{gsexp}) quoted in the body of the note. As seen there, this structure together with crossing symmetry is enough to fix all coefficients $\beta_{n,j}$. These values also agree with the ones found by explicit computations. 

\bigskip

\noindent
{\bf Twist conformal blocks.} 
The zeroth order correlator can be expressed in terms of twist conformal blocks $H^{(0)}_n(z,\bar z)$, defined as the contribution from operators with twist $\tau_n=4+2n$. 
\begin{equation}
\label{TCBdec}
\mathcal{H}^{(0)}(z,\bar z) = \sum_n  H^{(0)}_n(z,\bar z)
\end{equation}
The explicit form of (super-)conformal blocks leads to the following structure
\begin{equation}
\label{Hstructure}
H^{(0)}_n(z,\bar z) = \frac{(z \bar z)^{\tau_n/2}}{z-\bar z}\left(F_{\frac{\tau+2}{2}}(\bar z) h_n^{(0)}(z)-F_{\frac{\tau+2}{2}}( z) h_n^{(0)}(\bar z) \right)
\end{equation}
by plugging this structure into (\ref{TCBdec}) and expanding around $z=0$ the functions $h_n^{(0)}(z),h_n^{(0)}(\bar z)$ can be found
\begin{align*}
h_n^{(0)}(z)&= \rho_n\left( 2(3+n) \bar z ~_2F_1(3+n,4+n,2(3+n);z) \right.\\
&\left. +(3+n)(\bar z-2) ~_2F_1(4+n,4+n,2(3+n);z) \right)
\end{align*}
where $\rho_n=-\frac{\pi \Gamma (n+3)^2}{4^{2n+5} \Gamma \left(n+\frac{5}{2}\right) \Gamma \left(n+\frac{7}{2}\right)}$. Next we define the sequence of functions $H^{(m)}_n(z,\bar z)$ corresponding to extra insertions of $J^2=(\ell+n+3)(\ell+n+4)$
\begin{equation*}
H^{(m)}_n(z,\bar z)= \sum_{n,\ell} \frac{a^{(0)}_{n,\ell}}{J^{2m}} u^{2+n}  g_{n,\ell}(u,v)
\end{equation*}
$J^2$ is the eigenvalue of a specific quadratic Casimir operator. More precisely, $H^{(m)}_n(z,\bar z)$ admits the same factorisation as in (\ref{Hstructure}) with $h_n^{(0)}(z) \to h_n^{(m)}(z)$, where the functions $h_n^{(m)}(z)$ satisfy the following recursive relation
\begin{equation*}
{\cal D}_{su}  h_n^{(m+1)}(z) =h_n^{(m)}(z),~~~~{\cal D}_{su} = z^{-n-3} D z^{n+3}
\end{equation*}
and $D=(1-z)z^2\partial^2-z^2 \partial$. With this recursion relation together with the expression for $h_n^{(0)}(z)$ we can find $H^{(m)}_n(z,\bar z)$ exactly for the first few values of $m$ and also as various expansions. The divergent behaviour as $\bar z \to 1$ for $m=2,3,\cdots$ will be important for us. From the explicit answer we see
\begin{equation*}
\left. h_n^{(0)}(\bar z) \right|_{div}=\frac{a_n}{(1-\bar z)^2}+\frac{b_n}{1-\bar z}
\end{equation*}
It can be then seen that
\begin{equation*}
h_n^{(m)}(\bar z) = q_n^{(m)}(\bar z)  \log^2(1-\bar z),~~~m=2,3,\cdots
\end{equation*}
with $q_n^{(m)}(\bar z) \sim (1-\bar z)^{m-2}$ as $\bar z \to 1$ and
\begin{equation*}
{\cal D}_{su} q_n^{(m+1)}(\bar z) = q_n^{(m)}(\bar z) 
\end{equation*}
The functions $q_n^{(m)}(\bar z)$ can be build recursively to any desired order. 

\bibliography{sugraN4Notes}

\end{document}